\documentclass[epj]{webofc}
\usepackage[varg]{txfonts}   
\wocname{EPJ Web of Conferences}
\woctitle{CONF12}
\usepackage{color}
\usepackage{colordvi}

\usepackage[normalem]{ulem}

\usepackage{color}

\newcommand{\beqn}{\begin{eqnarray}}
\newcommand{\eeqn}{\end{eqnarray}}
\newcommand{\be}{\begin{equation}}
\newcommand{\ee}{\end{equation}}

\newcommand{\ba}{\begin{array}{c}}
\newcommand{\bat}{\begin{array}{cc}}
\newcommand{\ea}{\end{array}}

\newcommand{\bi}{\begin{itemize}}
\newcommand{\ei}{\end{itemize}}

\newcommand{\Frac}[2]{\frac{\displaystyle #1}{\displaystyle #2}}

\newcommand{\cO}{{\cal O}}

\newcommand{\mB}{\mathcal{B}}
\newcommand{\mC}{\mathcal{C}}

\newcommand{\mS}{\mathcal{S}}
\newcommand{\mT}{\mathcal{T}}

\newcommand{\lsim}{\stackrel{<}{_\sim}}
\newcommand{\gsim}{\stackrel{>}{_\sim}}

\newcommand{\Int}{\displaystyle{\int}}

\newcommand{\bear}{\begin{eqnarray}}
\newcommand{\eear}{\end{eqnarray}}
\newcommand{\nn}{\nonumber}

\newcommand{\tab}{\hspace*{0.5cm}}

\begin{document}
\selectlanguage{english}
\title{Sum-rule constraints on possible diphoton resonances at LHC}
%
%

\author{
P. Roig\inst{1}\fnsep\thanks{\email{proig@fis.cinvestav.mx}}
\and
J.J. Sanz-Cillero \inst{2}\fnsep\thanks{\email{jjsanzcillero@ucm.es}} \thanks{Speaker. 
Parallel talk at the XII Quark Confinement and the Hadron
Spectrum Conference (CONF12), Thessaloniki, Greece, 29 August – 3 September, 2016.}
}

\institute{
Departamento de F\'isica, Centro de Investigaci\'on y de Estudios Avanzados del Instituto Polit\'ecnico Nacional,
Apartado Postal 14-740, 07000 M\'exico City, M\'exico
\and
Departamento de F\'\i sica Te\'orica I, Universidad Complutense de Madrid, E-28040 Madrid, Spain
}

\abstract{%
The study of the forward scattering amplitude $V(k,\lambda) V(k'\lambda')\to V(k,\lambda) V(k',\lambda')$
of real massless gauge bosons $V$, e.g. photons or gluons, leads to a sum-rule that can be used to investigate
beyond the Standard Model signals at LHC in the $\gamma\gamma$ channel. The sum-rule only relies on general properties
such as analyticity, unitarity and crossing. We use the now buried ``750~GeV diphoton resonance'' as a case of study
to exemplify the constraints that the forward sum-rule requires to any new physics candidate. In the case of a large $\gamma\gamma$
or $gg$ partial width, of the order of 10~GeV in our 750~GeV analysis, one finds that an infinite tower of
states with spin $J_R=2$ and higher must be ultimately incorporated to the beyond Standard Model theory in order to fulfill
the sum-rule. We expect these techniques may be useful in next diphoton searches at LHC and future colliders.
}
\maketitle
\section{Introduction}
\label{intro}

On December 2015, the 13~TeV integrated luminosity
of 3.2~fb$^{-1}$ and 2.6~fb$^{-1}$ in ATLAS and CMS, respectively,
showed an excess in the diphoton spectrum,
hinting the presence of a resonance with invariant mass around 750~GeV~\cite{exp}.
The expectancy grew on March 2016, at {\it Rencontres de Moriond},
after reanalyses of ATLAS and CMS data from December 2015, together with
the recovery and study of additional 0.6~fb$^{-1}$ CMS (at 0~T)~\cite{Moriond}.
ATLAS showed the highest significance and pushed
the production cross section $\sigma(pp\to R \to \gamma\gamma)$ up,
with the combined value of $4.2\pm 2$~fb ($4.2\pm 2.6$~fb) for
the possible spin--0 (spin--2)
diphoton resonance at 750~GeV~\cite{Kim:2015ksf}.
The best fit preferred a broad total width of $\sim 45$~GeV~\cite{Kim:2015ksf}
and some works pointed out $\cO(GeV)$ partial widths
for $R\to gg$ and $R\to \gamma\gamma$~\cite{Csaki:2015vek,Csaki:2016raa},
though with huge uncertainties.
These features were difficult to  be fulfilled for
weakly-coupled beyond Standard Model (BSM) theories, where the assumed perturbativity
broke down around the TeV region~\cite{Staub:2016dxq,Bertuzzo:2016fmv}).
Thus, Strongly-coupled models and their higher dimensional duals
seemed to be favoured~\cite{Veneziano,RS,KK-graviton,Gouzevitch:2013qca,Sanz:2016auj}.

However, by August 2016, ATLAS and CMS accumulated 12.9~fb$^{-1}$
and 15.4~fb$^{-1}$ data, respectively,
showing essentially no significant excess around 750~GeV~\cite{ICHEP16}.
Even though the ``750~GeV diphoton resonance'' is now buried this does not mean
the end of searches in the diphoton channel. From this perpective, the theoretical constraints of
the analysis~\cite{Roig:2016tda} this talk is based on may be useful in future scans:
a lone scalar resonance with a large diphoton partial width leads to inconsistencies with
the basic assumptions of unitarity, crossing and analyticity. These issues can be only solved in two ways:
1) the
appearance of states with spin $J_R\geq 2$
(expected in composite BSM theories
or higher dimensional duals of strongly coupled field
theories)~\cite{Veneziano,RS,KK-graviton,Gouzevitch:2013qca,Sanz:2016auj};
2) the breakdown of perturbation theory around
the TeV~\cite{Staub:2016dxq,Bertuzzo:2016fmv}.

These proceedings are organized as follows:
In Sec.~\ref{sec:sum-rule} we derive the forward
sum-rule (FSR) that constrains the scattering of real spin--1 particles. We then
study  in Sec.~\ref{sec:scalar-SR} the impact of a scalar resonance, e.g. produced in the diphoton channel,
on the FSR.
One finds an unbalance in the FSR which requires the presence of additional contributions,
where a spin--2 resonance appears as a potential candidate.
We note that the sum-rule studied here only relies on unitarity, analyticity and crossing symmetry
and is therefore fulfilled in any possible BSM extension that assumes these general properties.
Finally in Sec.~\ref{sec:LHC} we provide some numerical estimates of what kind of $J_R=2$ resonance signal
one could expect if a spin--0 resonance shows up in LHC's diphoton spectrum, where we take the
``750~GeV resonance'' as a case of study to exemplify the analysis.
Some conclusions are gathered in Sec.~5.

\section{Theoretical framework: forward sum-rules for spin--1 particle scattering}
\label{sec:sum-rule}

Let us consider a spin--1 particle $V(k,\lambda)$ with momentum $k$
and helicity $\lambda=\pm 1$. The particle $V$ is assumed to be
described
by a real field in quantum field theory (QFT).  In principle it may carry other group indices,
but this will not be relevant for the derivation of the sum-rule,
only for some details of the later low-energy matching. Examples of this type of particles $V$ would be
the photon $\gamma$ or the gluon $g^a$
(for a fixed gluon colour index $a$, not averaged or summed).

Ref.~\cite{Roig:2016tda} studied the forward collision ($t=0$),
$T_{\Delta\lambda}$, as a function of the kinematical variable $\nu\equiv (s-u)/2$,
\footnote{
The use of the kinematical variable $\nu\equiv (s-u)/2$ is customary in fixed-$t$ dispersive analyses
of scattering amplitudes with definite $s\leftrightarrow u$ crossing properties (see in general Ref.~\cite{PWA}).
In particular, for $t=0$ one has $\nu=s$.
}
\bear
T_{\Delta\lambda}(\nu) =
T(V(k,\lambda)V(k',\lambda')\to V(k,\lambda)V(k',\lambda')) \, . \;\;
\eear
with helicity difference $\Delta \lambda=|\lambda-\lambda'|$.
For instance, the case $V=\gamma$ corresponds to the forward scattering
$\gamma(k,\lambda)\gamma(k'\lambda')\to\gamma(k,\lambda)\gamma(k'\lambda')$.

\begin{figure}[!t]
\centering
\includegraphics[width=5cm,clip]{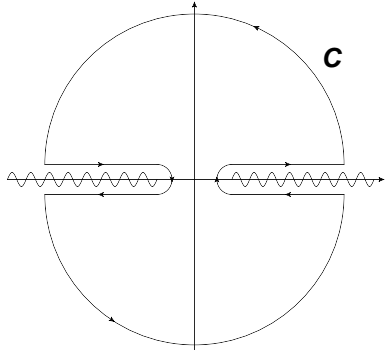}
\caption{Integration contour for the sum-rule complex integral.}
\label{fig:contour}
\end{figure}

We now proceed to impose the three basic ingredients of the sum-rule:
\begin{enumerate}

\item{\bf Analyticity:}
In Fig.~\ref{fig:contour} one can see the analytical structure of the function
$T_{\Delta\lambda}(\nu)$ in the complex $\nu$--plane. There
is a right-hand absorptive cut that corresponds
to the intermediate on-shell states when $s>0$ (s--channel cut)
and a left-hand cut related to the intermediate production of on-shell states when $u>0$
(u-channel cut).
This allows us to write down the Cauchy integral at the close contour $C$ in Fig.~\ref{fig:contour},
\bear
\Frac{1}{2\pi i}\oint_C {\rm d\nu '} \Frac{T_{\Delta \lambda}(\nu')}{\nu' (\nu' -\nu) }
\quad=\quad \Frac{T_{\Delta\lambda}(\nu)}{\nu} +\Frac{T_{\Delta\lambda}(0)}{(-\nu)} \, ,
\label{eq:Cauchy-int}
\eear
taking $\nu$ any complex value within the contour $C$.
In the right-hand side (rhs) we have used Cauchy's theorem and the fact that the in the first Riemann
sheet, where this integral is performed, $T_{\Delta\lambda}(\nu')$ is
analytical within the contour $C$
and the only poles are at $\nu'= 0$ and $\nu'=\nu$
due to the denominator in the integrand.

\item{\bf Unitarity:}
scattering cross sections are limited by
the Froissart bound~\cite{Froissart}. This implies constraints on the scattering amplitude
$T(V(k,\lambda)V(k'\lambda')\to X)$ for any final product $X$
and, in particular, on the forward scattering amplitude in the form
\bear
|T_{\Delta\lambda}(\nu)|^2 \quad < \quad \mC \, \nu\, \ln^2(\nu/\nu_0)\, ,
\eear
for some constants $\mC$, $\nu_0$. This means that the integral
in Eq.~(\ref{eq:Cauchy-int}) over the external circular parts of the contour $C$ vanishes
when its radius goes to infinity and the once-subtracted integrals on the real axis converge:
\bear
T_{\Delta\lambda}(\nu)- T_{\Delta\lambda}(0) &=&
 \Frac{\nu}{2\pi i}\oint_C {\rm d\nu '} \Frac{T_{\Delta \lambda}(\nu')}{\nu' (\nu' -\nu) }
\label{eq:SR-int1}
\\
&&
\hspace*{-3.25cm}
=  \Frac{\nu}{2\pi i}
\Int_{-\infty}^{\nu^{\rm (u)}_{\rm th}} {\rm d\nu'}
\Frac{( T_{\Delta \lambda}(\nu'+i\epsilon)- T_{\Delta \lambda}(\nu'-i\epsilon))}{\nu' (\nu' -\nu) }
+ \Frac{\nu}{2\pi i}
\Int_{\nu_{\rm th}^{\rm (s)}}^{\infty}  {\rm d\nu'} \Frac{( T_{\Delta \lambda}(\nu'+i\epsilon)- T_{\Delta \lambda}(\nu'-i\epsilon))}{\nu' (\nu' -\nu) }
\, ,
\nn
\eear
with $\nu_{\rm th}^{\rm (s)}$ and $\nu_{\rm th}^{\rm (u)}$ the threshold of the right-hand and left-hand cuts,
respectively. Both thresholds are related below through crossing symmetry.
By the analyticity Schwartz reflection principle one can rewrite
$T_{\Delta\lambda}(\nu'-i\epsilon)=T_{\Delta\lambda}(\nu'+i\epsilon)^*$:
\bear
T_{\Delta\lambda}(\nu)- T_{\Delta\lambda}(0) &=&
\Frac{\nu}{\pi}
\Int_{-\infty}^{\nu^{\rm (u)}_{\rm th}}  {\rm d\nu'}
\Frac{{\rm Im} T_{\Delta \lambda}(\nu'+i\epsilon)}{\nu' (\nu' -\nu) }
+ \Frac{\nu}{\pi}
\Int_{\nu_{\rm th}^{\rm (s)}}^{\infty}  {\rm d\nu'} \Frac{{\rm Im} T_{\Delta \lambda}(\nu'+i\epsilon)}{\nu' (\nu' -\nu) }
\, ,
\label{eq:SR-int2}
\eear
where we have used that 2$i$~Im$T_{\Delta\lambda}(\nu'+i\epsilon)=T_{\Delta\lambda}(\nu'+i\epsilon)-T_{\Delta\lambda}(\nu'+i\epsilon)^*$.

\item{\bf Crossing symmetry:}
Let us consider a forward amplitude with definite helicity, e.g, $T(V(k,+)V(k',+)\to V(k,+) V(k',+))$.
If one now exchanges the incoming and outgoing real vector $V$ with momentum $k'$ in the latter example,
crossing symmetry tells us that the previous amplitude coincides with that for
$T(V(k,+)V(-k',-)\to V(k,+) V(-k',-))$, which corresponds to the replacements $k'\to -k'$, $\lambda'\to -\lambda'$
and $\nu\to -\nu$. Hence, in general one has the relation
\bear
T_{\Delta\lambda}(\nu)&=& T_{\overline{\Delta\lambda}}(-\nu)\, ,
\eear
with $\overline{\Delta\lambda}=2-\lambda$.
The first implication of this crossing relation
is that the left-hand and right-hand thresholds are related in the form
$\nu^{\rm (s)}_{\rm th}=\, -\, \nu^{\rm (u)}_{\rm th}\equiv \nu_{\rm th}$.
This implies that in the previous integral~(\ref{eq:SR-int1})
one can rewrite the left-hand cut as a right-hand cut integral in the way
\bear
T_{\Delta\lambda}(\nu)- T_{\Delta\lambda}(0) &=&
\Frac{\nu}{\pi}
\Int_{\nu_{\rm th}}^{\infty}  {\rm d\nu'}
\left(
- \Frac{{\rm Im} T_{\overline{\Delta \lambda}}(\nu'+i\epsilon)}{\nu' (\nu' +\nu) }
+   \Frac{{\rm Im} T_{\Delta \lambda}(\nu'+i\epsilon)}{\nu' (\nu' -\nu) }\right)
\, .
\label{eq:SR-int3}
\eear

\end{enumerate}

This equation provides the master relation for the sum-rule studied in Ref.~\cite{Roig:2016tda}
and the Roy-Gerasimov-Moulin sum-rule for the inclusive
$\gamma\gamma$ cross section
${\sigma_{\Delta\lambda}=   \sigma(\gamma(k,\lambda)\gamma(k',\lambda')\to X)   }$~\cite{photon-SR}.
Its n--th derivative evaluated at $\nu=0$ yields
\bear
\Frac{1}{n!}\Frac{\rm d^n}{\rm d\nu^n} T_{\Delta\lambda}(0)  
&=&
\Frac{1}{\pi} \Int_{\nu_{\rm th}}^{\infty}  \Frac{{\rm d\nu'}}{\nu^{'\,\, n+1}}
\left( (-1)^{n} {\rm Im} T_{\overline{\Delta \lambda}}(\nu'+i\epsilon)
\, +\, {\rm Im} T_{\Delta \lambda}(\nu'+i\epsilon)  \right)
\, ,
\label{eq:n-der-SR}
\eear
for $n\geq 1$.

In the case when $V$ is a massless abelian gauge boson like, e.g., the photon,
the low-energy forward scattering is provided by the Euler-Heisenberg Lagrangian~\cite{EH-EFT}.
The effective field theory (EFT) operators contributing to this process
have dimesion~8 or higher and, therefore, the low-energy amplitude behaves at $\nu\to 0$ like
\bear
T_{\Delta\lambda}(\nu)\,\, \stackrel{\nu\to 0}{=}\,\,
c_{\Delta\lambda} \nu^2 \quad+ \quad \cO(\nu^3)\, .
\label{eq.EFT}
\eear
Thus, since $T_{\Delta \lambda}'(0)=0$, Eq.~(\ref{eq:n-der-SR}) for $n=1$
implies the forward sum-rule in~\cite{Roig:2016tda},
\bear
0 &=& \Frac{1}{\pi}\Int_{   \nu_{\rm th}      }^\infty \Frac{\rm d\nu'}{(\nu')^{2}}
\left( {\rm Im}T_2(\nu'+i\epsilon) -  {\rm Im}T_0 (\nu'+i\epsilon) \right) \, .
\label{eq.SR}
\eear
It is equivalent to the Roy-Gerasimov-Moulin
sum-rule for the inclusive $\gamma\gamma$ cross section
${    \sigma_{\Delta\lambda}=   \sigma(\gamma(k,\lambda)\gamma(k',\lambda')\to X)   }$~\cite{photon-SR}:
\bear
0 &=& \frac{1}{\pi}  \Int_{    \nu_{\rm th}    }^\infty \Frac{\rm d\nu'}{\nu'}
\left[ \sigma_2(\nu') -  \sigma_0 (\nu') \right] \, ,
\label{eq.RGM-SR}
\eear
by means of the relation
$\sigma_{\Delta\lambda}(\nu')={\rm Im}T_{\Delta\lambda}(\nu'+i\epsilon)/\nu'$.

Eq.~(\ref{eq:n-der-SR}) implies further sum-rules.
For instance, for even $n$ the right-hand side integral is positive-definite and one has
$d^nT_{\Delta\lambda}/d\nu^n >0$ at $\nu=0$.
In the (abelian gauge boson) photon scattering case this means that, for instance,
the corresponding combination $c_{\Delta\lambda}$ of
Euler-Heisenberg low-energy constants is positive.

An analogous result can be derived for non-abelian gauge boson scattering
like, e.g., the $g^a g^a\to g^a g^a$ process.
Around $\nu\sim 0$, massive states contribute to the forward scattering
of two same-colour gluons through EFT operators of dimension~8
(with the colour trace of four field-strength tensors $G^{\alpha\beta}$)
or higher.
Notice that we do not discuss the general $g^a g^b\to g^c g^d$ scattering;
all the initial and final gluons have the same colour $g^a$.
Although the pure Yang-Mills (YM) Lagrangian
yields no tree-level contribution to $g^a g^a\to g^a g^a$,
it generates a non-vanishing forward amplitude
$T_{\Delta \lambda}(\nu)^{\rm pure \, YM}$
which starts at the loop level. Nonetheless, the pure YM theory
is well behaved in the ultraviolet and fulfills Froissart's bound and
the once-subtracted dispersion relation~(\ref{eq:SR-int2}).
For this reason, in the case of non-abelian gauge boson scattering
(e.g., $g^a(k) g^a(k')\to g^a(k) g^a(k')$)
the replacement $T_{\Delta \lambda}(\nu) \longrightarrow
\widetilde{T}_{\Delta \lambda}(\nu)=[T_{\Delta \lambda}(\nu)-T_{\Delta \lambda}(\nu)^{\rm pure\, YM}]$
is implicitly assumed in the previous equations, leading
to a similar sum-rule (\ref{eq.SR}) for $\widetilde{T}_{\Delta \lambda}(\nu)$.
This result does not apply to the spontaneously broken gauge symmetry case,
as physics beyond the pure YM theory can generate a contribution
to the amplitude at $\nu\to 0$ and, hence,
to the l.h.s. of the sum-rule~(\ref{eq.SR}). From the EFT point of view
this means that there are gauge invariant operators of higher dimension
that contribute to the forward amplitude and its
derivative at $\nu=0$. In the case of the forward $ZZ$ scattering
in the Standard Model (SM), for instance,
$T_{\Delta\lambda}(0),\, T'_{\Delta\lambda}(0)\neq 0$ already at tree-level
due to the exchange of a massive Higgs.

The contribution from neutral colourless resonances $R$
to the spectral function is given at tree-level by
\bear
{\rm Im}T_{\Delta\lambda}(\nu+i\epsilon) \bigg|_R
&=&
\sum_R 16 \pi^2 \, (2 J_R+1)
\, M_R\, \Gamma_{R\to [VV]_{\Delta\lambda}}\,\, \delta(\nu-M_R^2) \, ,
\nn
\eear
and turns (\ref{eq.SR}) into
    \bear
0 &=& \sum_R\, 16\pi \, (2J_R+1) \,
\Frac{ ( \Gamma_{R\to [VV]_2}   - \Gamma_{R\to [VV]_0} ) }{M_R^3}
\qquad  \mbox{+ non-R,}
\label{eq.SR-narrow}
\eear
where
\bear
\hspace*{-0.75cm}
\Gamma_{R\to [VV]_0}=\Gamma_{R\to V(+)V(+)}+\Gamma_{R\to V(-)V(-)}\, ,
\,\,
\Gamma_{R\to [VV]_2}=\Gamma_{R\to V(+)V(-)}\, ,
\,\,
\Gamma_{R\to VV}= \Gamma_{R\to [VV]_0}+ \Gamma_{R\to [VV]_2}\, .
\eear
The non-R contribution represents the loop diagrams in $VV\to VV$ without
an intermediate $s$ and $u$--channel resonance at tree-level.

The importance of these sum-rules relies on the fact that when
the vector $V$ is massless
the lowest-spin resonances ($J_R=0$) can only decay into
$V(k,\lambda)V(k',\lambda')$ pairs with $\Delta\lambda=0$.
Hence, these spin--0 resonances give a negative contribution
to the sum-rule (\ref{eq.SR-narrow}). The positive terms with $\Gamma_{R\to [VV]_2}$
only appear for higher spin resonances
with $J_R\geq 2$~\cite{photon-SR,Pascalutsa:2010,Pascalutsa:2012,Panico:2016}:
\bear
\hspace*{-0.75cm}\Gamma_{R\to VV}= \Gamma_{R\to [VV]_0}\geq 0\, , \,\,\, \Gamma_{R\to [VV]_2}=0\quad
\mbox{ for } J_R=0
\tab&\longrightarrow&\tab
(\Gamma_{R\to [VV]_2}-\Gamma_{R\to [VV]_0}) \leq  0\, ,
\nn\\
\hspace*{-0.75cm}\Gamma_{R\to VV}= \Gamma_{R\to [VV]_0}=\Gamma_{R\to [VV]_2}=0\quad
\mbox{ for } J_R=1
\tab&\longrightarrow&\tab
(\Gamma_{R\to [VV]_2}-\Gamma_{R\to [VV]_0}) =0\, ,
\nn\\
\hspace*{-0.75cm}\Gamma_{R\to [VV]_0}\geq 0\, , \quad \Gamma_{R\to [VV]_2}\geq 0\quad
\mbox{ for } J_R=2,4,6...
\tab&\longrightarrow&\tab
(\Gamma_{R\to [VV]_2}-\Gamma_{R\to [VV]_0}) \gtreqless 0\, ,
\nn\\
\hspace*{-0.75cm}\Gamma_{R\to VV}= \Gamma_{R\to [VV]_2}\geq 0\, , \,\,\, \Gamma_{R\to [VV]_0}=0\,\,\,
\mbox{ for } J_R=3,5,7...
\tab&\longrightarrow&\tab
(\Gamma_{R\to [VV]_2}-\Gamma_{R\to [VV]_0})\geq 0\, ,
\nn
\eear
On-shell resonances with $J_R=1$ are forbidden by the Landau-Yang theorem~\cite{Landau-Yang}
and those with $J_R=2,4,6...$ can in principle decay
into $[VV]_0$ and $[VV]_2$ states~\cite{Panico:2016}.
In QCD the $\gamma\gamma$ decay of the lowest-lying spin--2
resonances ($\mT=a_2,f_2,f_2'$)
predominantly occurs with helicity $\Delta\lambda=2$~\cite{Tensor-Helicity-2},
i.e., $\Gamma_{\mT \to \gamma\gamma}\approx  \Gamma_{\mT\to [\gamma\gamma]_2}$.
The sum-rule~(\ref{eq.SR}) is mostly saturated by
the lightest $J_R=0$ ($\pi^0,\eta,\eta'$)
and $J_R=2$ ($a_2,f_2,f_2'$)
meson multiplets~\cite{Pascalutsa:2010,Pascalutsa:2012}:
the large spin--2 positive contribution cancels out to a large extent
the large negative spin--0 contribution.
Similar thing happens in the case of (spin-2) massive
gravitons $G$~\cite{RS,KK-graviton,Gouzevitch:2013qca,Sanz:2016auj},
where the decay $G\to V(\lambda)V(\lambda')$
always occurs with $\Delta\lambda=2$ as the graviton couples
to the stress-energy tensor of the gauge field $V=\gamma, g^a$.

\section{Sum-rule with a large-width scalar: you can't just put it alone}
\label{sec:scalar-SR}

In this and next Sections, we will explore the FSR constraints on
any possible BSM physics that shows up in the $\gamma\gamma$ channel
at LHC or future colliders.
Let us assume the existence of a diphoton scalar (or pseudoscalar)
resonance $\mS$ similar to the now discarded 750~GeV candidate~\cite{exp,Moriond,ICHEP16}.
We will take the latter as a case of study
to exemplify the analysis of future experimental signals.

Conveniently reordering the sum-rule~(\ref{eq.SR-narrow}) one has
\bear
16\pi  \, \Frac{\Gamma_{\mS\to VV}}{M_{\mS}^3}
&=&    \sum_{R\neq \mS} \, 16\pi \, (2J_R+1) \,
\Frac{ ( \Gamma_{R\to [VV]_2}   - \Gamma_{R\to [VV]_0} ) }{M_R^3}
 \qquad \mbox{+ non-R.}
\label{eq.scalar-SR}
\eear
In the case of a large $\Gamma_{\mS\to VV}$ partial width,
resonances with $\Gamma_{R\to [VV]_2}\neq 0$ are needed
on the r.h.s. to fulfill the identity~(\ref{eq.scalar-SR}),
i.e., resonances with spin $J_R\geq 2$.
This implies the existence of a infinite tower of higher spin resonances
{\it \`a la Regge}~\cite{Veneziano} to cure the divergent high-energy behaviour
of the cross channel resonance exchanges in the partial wave amplitudes~\cite{PWA}.
These BSM resonance theory is non-renormalizable
and dual to an underlying strongly coupled theory where the resonances
are composite states.

This argumentation relies to a large extent on the fact
that the $\Gamma_{\mS\to VV}$ partial decay width is large, with
\bear
16\pi \Frac{\Gamma_{\mS\to VV} }{M_{\mS}^3}
\quad \sim \quad 1\, \mbox{TeV}^{-2}\, ,
\eear
as it occurred with the former scalar candidate with mass $M_{\mS}\sim 750$~GeV
and a partial width $\Gamma_{\mS\to VV} \sim 10$~GeV (for $V=\gamma,g^a$).
In this situation size matters, as we find very difficult
that the FSR~(\ref{eq.scalar-SR})
can be compensated by the non-resonant loop contributions to
$V(k)V(k')\to V(k) V(k')$.
Based on naive dimensional analysis, these
are found to be small,
of the order of
\bear
&&\Frac{1}{\pi}\Int_{\nu_{\rm th}}^\infty \, \, \Frac{\rm d\nu'}{(\nu')^{2}}  \,\,
{\rm Im}T_{\Delta\lambda}(\nu'+i\epsilon) \bigg|_{\rm non-R}
  \sim  \Frac{\alpha^2}{\nu_{\rm th}}\quad\sim
\quad 10^{-4}\, \mbox{TeV}^{-2}\, ,
\eear
 where, for possible new physics states in the non-R loop
in an underlying weakly interacting theory (if any),
we expect the thresholds to be $\nu_{\rm th}> (750$~GeV$)^2$.
More precisely, in Quantum Electrodynamics (QED) with either a scalar or
a spin--$\frac{1}{2}$ particle with charge $Q=1$,
the $\gamma\gamma$ cross-section difference
reaches a sharp global minimum with
$\sigma_2(\nu)-\sigma_0(\nu)
\gsim  - 8\alpha^2/\nu_{\rm th}$
right after the production threshold
due to the negative $\Delta\lambda=0$ contribution, then a wider global maximum
with $\sigma_{\Delta\lambda}(\nu)
\lsim 2\alpha^2/\nu_{\rm th}$
due to positive $\Delta\lambda=2$ production,
and finally a converging $1/\nu$ tail~\cite{Pascalutsa:2012}.
Thus, one finds that the pure QED one-loop amplitude
for $\gamma\gamma\to\gamma\gamma$ fulfills the sum-rule~(\ref{eq.SR}) on
its own and yields no correction to our FSR~\cite{Pascalutsa:2012}. Therefore,
in order to get a contribution from these background loops
to cancel the scalar resonance one
in Eq.~(\ref{eq.scalar-SR}), one should incorporate effects
beyond QED, which first enter at two loops. We find very unlikely
that these corrections are large enough to achieve this goal without
entering
a non-perturbative regime.

In the case of a small partial width $\Gamma_{\mS\to VV}$,
one does not need to include resonances with $J_R\geq 2$ and
the sum-rule can be fulfilled through the non-resonant loop terms in~(\ref{eq.scalar-SR}).
In that situation, the underlying theory is perturbative.
For instance, in the SM, the Higgs exchange yields a much smaller
contribution $16\pi \Gamma_{h\to\gamma\gamma}/m_h^3  \simeq 2.4\times 10^{-4} \, \mbox{TeV}^{-2}$~\cite{pdg},
which is cancelled out by the $\gamma\gamma\to\gamma\gamma$ loop amplitude
without any need of BSM physics.
For radiatively generated $\mS\to VV$ decays ($VV=\gamma\gamma, gg$)
in weakly interacting theories in the TeV, small partial widths are expected:
one would have $\Gamma_{\mS\to gg}\lsim  \alpha_S^2 M_\mS^3/(72\pi^3 v^2) \sim  10^{-5}$~TeV
for a Higgs-like decay $\mS\to gg$~\cite{Wilczek:1977zn,Djouadi:2005gi},
yielding a contribution to~(\ref{eq.SR-narrow}) of the order of
$16\pi  \Gamma_{\mS\to gg}/M_{\mS}^3 \lsim  2 \alpha_S^2 /(9\pi^2 v^2) \sim  10^{-3}$~TeV$^{-2}$.
These numbers get roughly two orders of magnitude smaller for $\mS\to \gamma\gamma$,
for a similar Higgs-like radiatively generated decay~\cite{Djouadi:2005gi}.
A perturbative BSM theory including a scalar resonance with a loop-induced decay similar to the SM one
needs either a large number of particles running in the intermediate
loop or huge hypercharges~\cite{Franceschini:2015kwy} to give a contribution to~(\ref{eq.scalar-SR})
of the order of 1~TeV$^{-2}$.
Achieving the latter and the required large background
non-resonant contribution to the sum-rule implies a departure from perturbativity in the TeV range,
where the BSM theory would enter a strongly coupled regime
(see e.g. Ref.~\cite{Bertuzzo:2016fmv}).
Thus, one would expect to have composite states of any total angular momenta $J_R\geq 2$ lying
in the non-perturbative energy range,
as nothing forbids excitations with an arbitrary orbital momentum, similar to what one observes in QCD.

\section{$J_R=0$ and $J_R=2$ resonances and $\gamma\gamma$ production at LHC}
\label{sec:LHC}

Let us assume that the lightest and lowest-spin resonances dominate
the FSR  sum-rule.
This means that the lightest scalar $\mS$
and tensor $\mT$ resonance partial widths
are related in the approximate form
\bear
\Frac{16\pi \Gamma_{\mS\to VV}}{M_{\mS}^3} \quad
\approx \quad \Frac{80\pi (\Gamma_{\mT\to [VV]_2} - \Gamma_{\mT\to [VV]_0}) }{M_{\mT}^3}
\quad \leq \quad \Frac{80\pi \Gamma_{\mT\to VV}}{M_{\mT}^3} \, .
\label{eq.S-T-rel}
\eear
If $\mT$ only decays into helicity $\Delta\lambda=2$ states
--as it happens in the case of massive gravitons~\cite{RS,KK-graviton,Gouzevitch:2013qca,Sanz:2016auj}--
Eq.~(\ref{eq.S-T-rel}) turns into an identity and one obtains
the lowest possible bound for $\Gamma_{\mT\to VV}$. We will assume
this lowest signal scenario
$\Gamma_{\mT\to VV}=\Gamma_{\mT\to [VV]_2}$ from here on and assume the relation
\bear
\Gamma_{\mT\to VV } & \approx & \Frac{\Gamma_{\mS\to VV}}{5}
\left( \Frac{M_\mT}{M_\mS}\right)^3 \, .
\label{eq.GT-GS-rel}
\eear

To illustrate how this relation may provide useful information
about the diphoton data we will exemplify the analysis
with the former 750~GeV candidate being our scalar $\mS$ and the lightest
tensor $\mT$ being provided by the 2.8~$\sigma$ (2.4~$\sigma$)
excess at $M_\mT\approx $ 1.6~TeV from ATLAS data presented in Moriond 2016~\cite{Moriond}
(in August 2016 in ICHEP~\cite{ICHEP16}).
Using these numbers and a large partial width
$\Gamma_{\mS\to VV}\approx 10$~GeV as reference values one obtains
\bear
\Gamma_{\mT\to VV } \quad &\approx & 20 \, \mbox{GeV}
\quad \left( \Frac{\Gamma_{\mS\to VV}}{10\, \mbox{GeV}}\right)
\quad \left(\Frac{0.75\, \mbox{TeV}}{M_\mS}\right)^3
\quad \left( \Frac{M_\mT}{1.6\, \mbox{TeV}}\right)^3 \, .
\eear
Assuming the inverted case $M_\mT\approx 750$~GeV and $M_\mS\approx 1.6$~TeV,
with $\Gamma_{\mT(750)\to VV}\sim  10$~GeV
would have led to a huge partial width
$\Gamma_{\mS(1600)\to VV}\sim  0.5$~TeV and a far too large and broad signal.

One can make a final exercise with our illustrative study with
$M_\mS=750$~GeV and $M_\mT=1.6$~TeV. Early 2016 combined analyses
of the LHC13 diphoton data yielded a cross section
$\sigma(pp\to \mS(750)\to \gamma\gamma)=(4.2\pm 2)$~fb~\cite{Kim:2015ksf},
with ATLAS13 data pushing for higher values and CMS13
for a lower cross section. The best fit to the 3.2~fb$^{-1}$ 2015 ATLAS13
(the 2.7~fb$^{-1}$ CMS13 data) pointed out an expected excess of
6.6~signal events~(8 signal events)~\cite{Kim:2015ksf}.
If $gg$ was the main was the main production channels
one had
\bear
\sigma(pp\to R ) =
\Frac{(2J_R+1) C_{gg}(M_R)\, \Gamma_{R\to gg} }{
s M_R} \, ,
\qquad
\sigma(pp\to R \to\gamma\gamma)
= \sigma(pp\to R)\, \mB_{R\to \gamma\gamma} \, ,
\eear
with $\sqrt{s}=8\ (13)$~TeV~\cite{Kim:2015ksf,Franceschini:2015kwy}.
Adding the $\gamma\gamma$ fusion~\cite{Csaki:2015vek,Csaki:2016raa}
contribution does not change this picture.
In the case of lepton colliders the latter channel might be
the dominant production mode and one should make then the replacement
$C_{gg}(M_R)$ and
$C_{gg}(M_R) \Gamma_{R\to gg}$ by
$C_{\gamma\gamma}(M_R)$ and
$C_{\gamma\gamma}(M_R) \Gamma_{R\to \gamma\gamma}$, respectively,
in the expressions for the cross sections
and the tensor-to-scalar ratios in this Section.
The different partonic contributions ($C_{gg}$, $C_{uu}$, ...)
can be found in Ref.~\cite{Franceschini:2015kwy},
also including their (parton-dependent) scaling between 8 and 13 TeV
collision energies corresponding to LHC run-I and run-II data, respectively.
Larger energies and higher-order corrections increase the relative importance
of the $gg$ contribution.
The ratio of tensor-to-scalar
production cross section is given by
\bear
\Frac{\sigma(pp\to \mT)}{\sigma(pp\to \mS)}
\quad =\quad
\Frac{5 C_{gg}(M_\mT)}{C_{gg}(M_\mS)} \,
\Frac{ \Gamma_{\mT\to gg} /M_\mT }{
\Gamma_{\mS\to gg} /M_\mS }
\quad \approx \quad  \Frac{C_{gg}(M_\mT)}{C_{gg}(M_\mS)}
\,\left(\Frac{M_\mT}{M_\mS}\right)^2\,
\, .
\label{eq.R-prod}
\eear
For the values $M_{\mS}=750$~GeV and $M_{\mT}=1.6$~TeV,
an estimate through MG5\_aMC~\cite{Madgraph}
with the parton distribution function set
NN23LO1~\cite{pdf} yields
\bear
\Frac{C_{gg}(M_\mT)}{C_{gg}(M_\mS)} \left(\Frac{M_\mT}{M_\mS}\right)^2
\, =\, 7.0\, \%  \, ,
\label{eq.pdf}
\eear
for $\sqrt{s}=13$~TeV
and roughly a factor of two smaller for $\sqrt{s}=8$~TeV.
Assuming that $\mT\to \gamma\gamma$ and $\mS\to\gamma\gamma$ decays have
a similar branching ratio --which is implied by~(\ref{eq.GT-GS-rel})
when $\gamma\gamma$ and $gg$ are the main decay channels--
one expects the tensor-to-scalar ratio
\bear
\Frac{\sigma(pp\to \mT)}{\sigma(pp\to \mS)}
\quad \approx \quad \Frac{0.29\, \mbox{fb}}{4.2\, \mbox{fb}}\,
\, .
\label{eq.R-prod}
\eear
Considering a detection efficiency for $\mT(1600)$ similar
to that obtained for the $\mS(750)$ one
it is possible to perform a rough estimate of the number of resonant diphoton
events $N_{pp\to \mT\to\gamma\gamma}$ that should have been observed for an integrated luminosity of
15~fb$^{-1}$ at each detector:
\bear
N_{pp\to \mT\to\gamma\gamma}\approx 2.2\, \mbox{signal events (ATLAS 13)}\, ,
\qquad
N_{pp\to \mT\to\gamma\gamma}\approx 3.1\, \mbox{signal events (CMS 13)}\, ,
\eear
with the SM background being much smaller than
these numbers  at 1.6~TeV.
Thus, joining both experiments, we should have observed around 5 events
by August 2016. Unfortunately, at the same time that the significance
of the 750~GeV resonance faded away no clear signal could be observed
for 1.6~TeV diphoton excess.

\section{Conclusions}

In this talk we have presented the work~\cite{Roig:2016tda},
where we showed that a colourless neutral scalar with a large $gg$ or $\gamma\gamma$
partial width cannot show up alone at LHC or future colliders.
Our FSR analysis implies that either
an infinite tower of resonance with spin $J_R\geq 2$ {\it \`a la Regge}
must be present in the underlying theory,
or this enters a non-perturbative
regime at the TeV --triggering the appearance of composite states
and resulting in similar conclusions--.

Even though the 750~GeV diphoton resonance has essentially disappeared
back into the background after August 2016 LHC13 data,
we have exemplified in these proceedings the kind of constraints that
it is possible to extract with our FSR.
Spin--2 (or higher) resonances are required to cancel out the scalar contribution to the sum-rule
in the case of large partial width $\Gamma_{\mS\to VV}\sim 10$~GeV.
Taking, the other higher significant excess in the data at 1.6~TeV
as our tensor candidate we were able to perform an estimate
of how the $\mT(1600)$ excess should also appear in following runs.

This powerful model-independent approach is based on
basic principles such as analyticity,
unitarity and crossing symmetry and can be easily applied in future
diphoton analyses in case a new resonant signal appears. It may serve to
pin down the necessary accompanying BSM states in the case of
large partial widths $\Gamma_{R\to VV}$ which cannot be compensated
in the FSR through perturbative loop contributions of a weakly interacting BSM
theory.

Even in the case of small partial-width resonance, the FSR may prove
to be useful, as the possible resonance excess generates
an unbalance in the sum-rule that must be compensated. However, in that
case,in addition to further resonances one should look for an excess
in the background as the sum-rule may now be compensated through
the non-resonant loop diagrams of a weakly-interacting BSM theory.
This task may be more cumbersome but still worthy to be analyzed
in order to confirm or discard any possible new diphoton resonance.

\section*{Acknowledgements}

PR acknowledges funding from Conacyt, M\'exico,
through projects 296 ('Fronteras de la Ciencia')
, 236394, 250628 ('Ciencia B\'asica')
and SNI.
The work of JJSC was supported by the Spanish Ministry MINECO
under grant
FPA2013-44773-P                         
and  MINECO:FPA2014-53375-C2-1-P.       

\end{document}